\documentclass[manuscript]{aastex}

\usepackage{rotating}
\usepackage{tikz}
\usepackage{amsmath}
\usepackage{amssymb}
\usepackage{hyperref}

\slugcomment{Ap.J., 847, L19 (2017 Sep 28)}

\shorttitle{Comet C/2017 K2 At Large Heliocentric Distance}
\shortauthors{Jewitt}

\begin{document}

\title{A Comet Active Beyond the Crystallization Zone}

%
%

\author{David Jewitt$^{1,2}$, Man-To Hui$^1$,  Max Mutchler$^3$, Harold Weaver$^4$, Jing Li$^{1}$, Jessica Agarwal$^5$ 
}

\affil{$^1$Department of Earth, Planetary and Space Sciences,
UCLA, 
595 Charles Young Drive East, 
Los Angeles, CA 90095-1567\\
$^2$Department of Physics and Astronomy,
University of California at Los Angeles, \\
430 Portola Plaza, Box 951547,
Los Angeles, CA 90095-1547\\
$^3$ Space Telescope Science Institute, 3700 San Martin Drive, Baltimore, MD 21218 \\
$^4$ The Johns Hopkins University Applied Physics Laboratory, 11100 Johns Hopkins Road, Laurel, Maryland 20723  \\
$^5$ Max Planck Institute for Solar System Research, Justus-von-Liebig-Weg 3, 37077 G\"ottingen, Germany\\
}

\email{jewitt@ucla.edu}


\begin{abstract}

We present observations showing in-bound long-period comet C/2017 K2 (PANSTARRS) to be active at record  heliocentric distance. Nucleus temperatures are  too low (60 K to 70  K) either for water ice to sublimate or for amorphous ice to crystallize, requiring another source for the observed activity.  Using the Hubble Space Telescope we find a sharply-bounded, circularly symmetric dust coma 10$^5$ km in radius, with a total scattering cross section of $\sim$10$^5$ km$^2$.  The coma has a logarithmic surface brightness gradient  -1 over much of its surface, indicating sustained, steady-state  dust production.  A lack  of clear evidence for the action of  solar radiation pressure suggests that the dust particles are large, with a mean  size  $\gtrsim$ 0.1 mm.  Using a coma convolution model, we find a limit to the apparent magnitude of the nucleus $V >$  25.2 (absolute magnitude $H >$ 12.9). With assumed geometric albedo $p_V$ = 0.04, the limit to the nucleus  circular equivalent radius is $<$ 9 km.  Pre-discovery observations from 2013 show that the comet was also active at 23.7 AU heliocentric distance.  While neither water ice sublimation nor exothermic crystallization can account for the observed distant activity, the measured properties are consistent with activity driven by sublimating supervolatile ices such as CO$_2$, CO, O$_2$ and N$_2$.  Survival of supervolatiles at the nucleus surface is likely a result of the comet's recent arrival from the frigid Oort cloud.
\end{abstract}


\section{Introduction}

The comets are icy leftovers from planetary accretion, and are widely believed to be  compositionally the most pristine objects in the solar system.  They have survived since formation 4.6 Gyr ago in the Kuiper belt and Oort cloud  reservoirs, at temperatures below $\sim$40 K and $\sim$10 K, respectively.  

Most known comets are active only when inside the orbit of Jupiter, where  sublimation of the most abundant cometary volatile (water ice) is responsible  (Whipple 1950).   However, activity is occasionally observed in more distant comets (Jewitt 2009, Kulyk et al.~2016, Meech et al. 2009, Meech et al.~2017, Womack et al. 2017) for which numerous explanations have been proposed, most notably the exothermic crystallization of amorphous ice (Prialnik and Bar-Nun 1992). Other suggestions include reactions of unstable radicals created by prolonged cosmic ray bombardment (Donn and Urey 1956), polymerization reactions (Rettig et al.~1992), impact (Sekanina 1973), exothermic heat of solution (Miles 2016) and the sublimation of supervolatile ices (Womack et al.~2017).  

In comets observed outbound from perihelion, distant activity has a mundane explanation in terms of the slow propagation of heat acquired at perihelion and conducted into the  nucleus interior.  For example, the outbound Comet 1P/Halley (perihelion 0.6 AU) experienced an outburst at 14 AU which was readily explained in this way (Prialnik and Bar-Nun 1992), as was activity in outbound comet C/Hale-Bopp at 25 AU (Szabo et al. 2011). On the other hand, distant activity in an \textit{inbound} long-period comet cannot be explained by slow conduction, since the comet is approaching the planetary region from larger distances where  lower, not higher, temperatures prevail.  

Comet C/2017 K2 (PANSTARRS) (hereafter K2) was discovered on UT 2017 May 21 (Wainscoat et al.~2017).  Its orbit (which, as of 2017 July 24, is a hyperbola with perihelion 1.811 AU, semimajor axis -7231 AU, eccentricity e = 1.00025, and  inclination i = 87.6 deg) identifies K2  as  long-period comet (LPC), as does the small Tisserand parameter measured with respect to Jupiter ($T_J \sim$ 0), c.f.~Levison (1996).   Perihelion is expected on 2022 December 21.  As with other hyperbolic orbit comets, K2 is probably not of interstellar origin, but has been slightly deflected from a bound orbit by planetary perturbations or outgassing forces (Rickman 2014, Dones et al. 2015)\footnote{Corrected for planetary perturbations, the pre-entry orbit is parabolic with semimajor axis 12,800 AU, classifying K2 as dynamically new. See \url{http://www.oaa.gr.jp/~oaacs/nk/nk3387.htm} by Syuichi Nakano}.  We may thus infer that the surface of the nucleus of K2 is  warming from very low temperatures (as small  as $\sim$10 K in the Oort cloud) to the current 60 K or 70 K, triggering the observed activity.  The initial observations of K2 described here give us an exceptional opportunity to study an Oort cloud comet as it enters the planetary region.  

\section{Observations}
We secured six images each of 285 s duration using the UVIS mode of the WFC3 camera, under Hubble Space Telescope (HST) observing program GO 14939.  The WFC3 instrument contains two 2k$\times$4k charge-coupled device detectors with pixels 0.04\arcsec~on a side, providing a 162\arcsec$\times$162\arcsec~field of view from which we read out a 2k subarray (80\arcsec$\times$80\arcsec~field).  In order to secure maximum signal-to-noise ratio data, we used the extremely broad-band F350LP filter: this filter has central wavelength $\lambda_C =$ 6230\AA~when used on a sun-like spectrum and has a full-width at half-maximum (FWHM) of 4758\AA.  The six images were obtained in two groups of three, dithered on the CCD in order to provide protection against defective pixels.  The  images were shifted into alignment and combined into a single image of higher signal-to-noise ratio for analysis.

The resulting apparent magnitudes, $V$, were converted to absolute magnitudes using 

\begin{equation}
H = V - 5\log_{10}(r_H \Delta) - g(\alpha)
\label{abs}
\end{equation}

\noindent in which $r_H$ and $\Delta$ are the heliocentric and geocentric distances, respectively, and $g(\alpha)$ is a measure of the phase darkening at phase angle $\alpha$.  The phase coefficient is unmeasured; we assume $g(\alpha) = 0.04 \alpha$ based on observations of other active comets (Meech and Jewitt 1987, Jewitt and Meech 1987a) but note that the value of $g(\alpha)$ is not critical because the phase angles are small (Table \ref{geometry}).  

The effective cross-section for scattering was then calculated from 

\begin{equation}
C_d = \frac{1.5\times10^6}{p_V} 10^{-0.4 H}
\label{area}
\end{equation}

\noindent where $C_d$ is in km$^2$ and $p_V$ = 0.04 is our assumed value of the V-band geometric albedo.  The cross-sections within a set of fixed linear apertures are listed in Table (\ref{photometry}).  

Pre-discovery observations from the Canada-France-Hawaii 3.6 m telescope (CFHT) atop Mauna Kea were identified using the Canadian Astronomy Data Centre archive.  The comet was detected in four images taken using the MegaCam prime focus imager (0.185\arcsec~per pixel) on UT 2013 May 12 and one from May 13.  Images from the former date, each of 600 s integration through a U filter ($\lambda_C =$ 3743\AA, FWHM = 758\AA),  were shifted according to the motion of the comet and combined into a single 3000 s equivalent composite (Figure \ref{image}).  The point-spread function measured from stars had FWHM = 0.73\arcsec.  While the image of K2 is clearly non-stellar,  with FWHM = 1.5\arcsec~$\pm$ 0.2\arcsec, we were unable to meaningfully determine the surface brightness profile in detail owing to the low signal-to-noise ratio of the CFHT data.  The night was photometric to within a few percent.  Accordingly, we measured the magnitude of K2 using standard Megacam calibrations but checked the result using photometry of nearby field stars from the USNO-B catalog.  Within a $\theta$ = 2.3\arcsec~radius aperture (linear radius 40,000 km at the comet) we obtained U = 23.7$\pm$0.3.  We converted to V using the average color of comets U-V = 1.15 (derived from Solontoi et al.~2012 and Jewitt 2015), finding V = 22.5$\pm$0.5, where the error bar reflects both noise in the data and our best estimate of the possible uncertainty in the color of K2.  The corresponding absolute magnitude computed from Equation (\ref{abs}) is $H$ = 8.6$\pm$0.5, about 1.4$\pm$0.5 magnitude fainter than the 40,000 km measurement from 2017 (Table \ref{photometry}), albeit with considerable uncertainty.  

\section{Discussion}

\begin{figure}
\epsscale{0.99}
\plotone{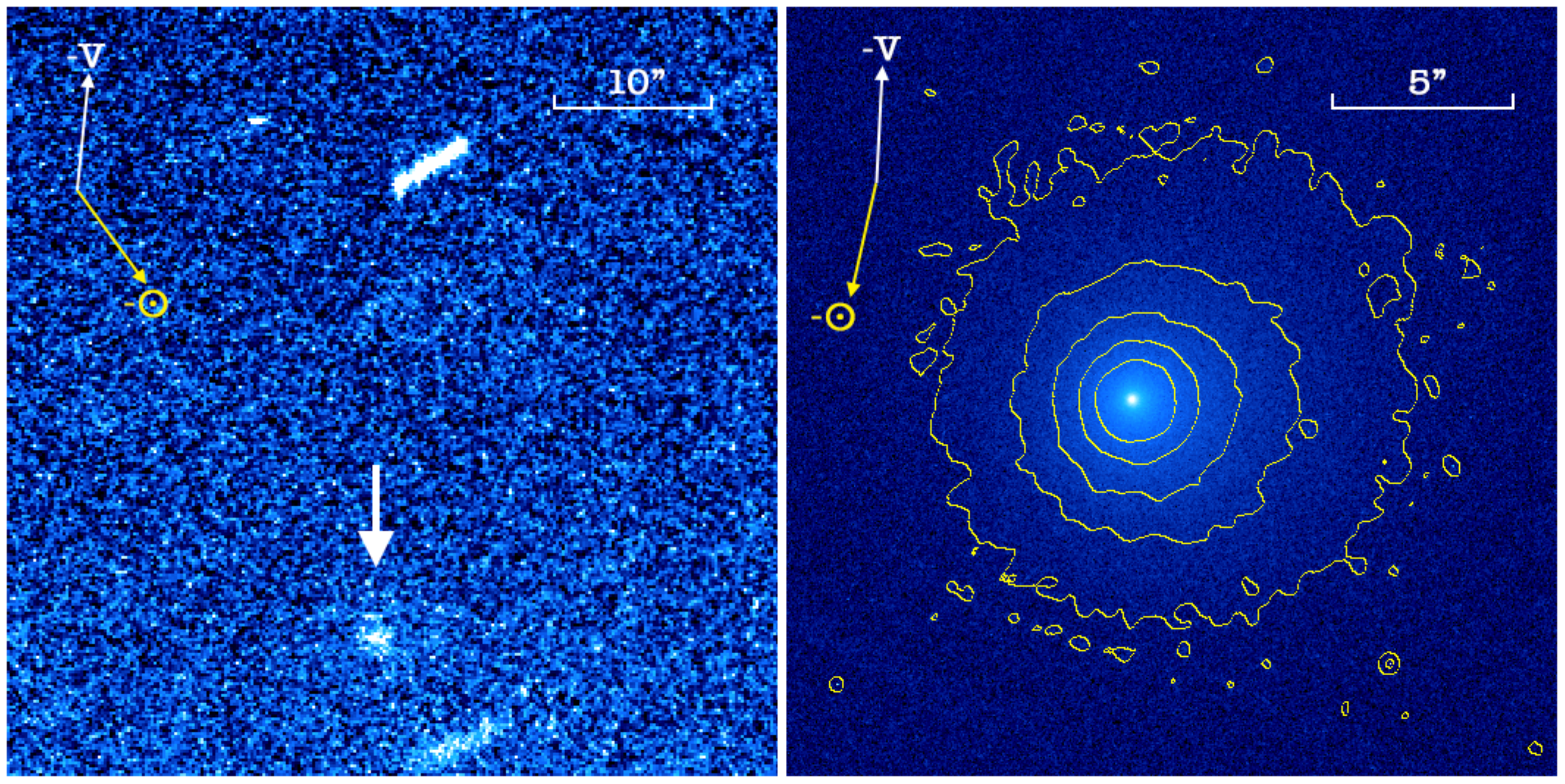}
\caption{Left: Prediscovery  CFHT image of C/2017 K2 (arrow) from UT 2013 May 12 at 23.765 AU. Right: HST image from UT 2017 Jun 27 at 15.874 AU.  The antisolar (-$\odot$) and negative velocity ($-V$) vectors are marked.   Both images have North to the top, East to the Left.
\label{image}}
\end{figure}

The central $\sim$5\arcsec~of the coma appears  nearly circularly symmetric in the plane of the sky (Figure \ref{image}).  A mild asymmetry  at larger radii, towards position angle 300\degr $\pm$10\degr, is aligned with neither the antisolar direction nor the projected orbit.  It presumably reflects a weak anisotropy in the ejection from the nucleus.   We searched for near-nucleus azimuthal structures (``jets'') by subtracting annular average brightnesses from the data, but found none.  

\begin{figure}
\epsscale{.90}
\plotone{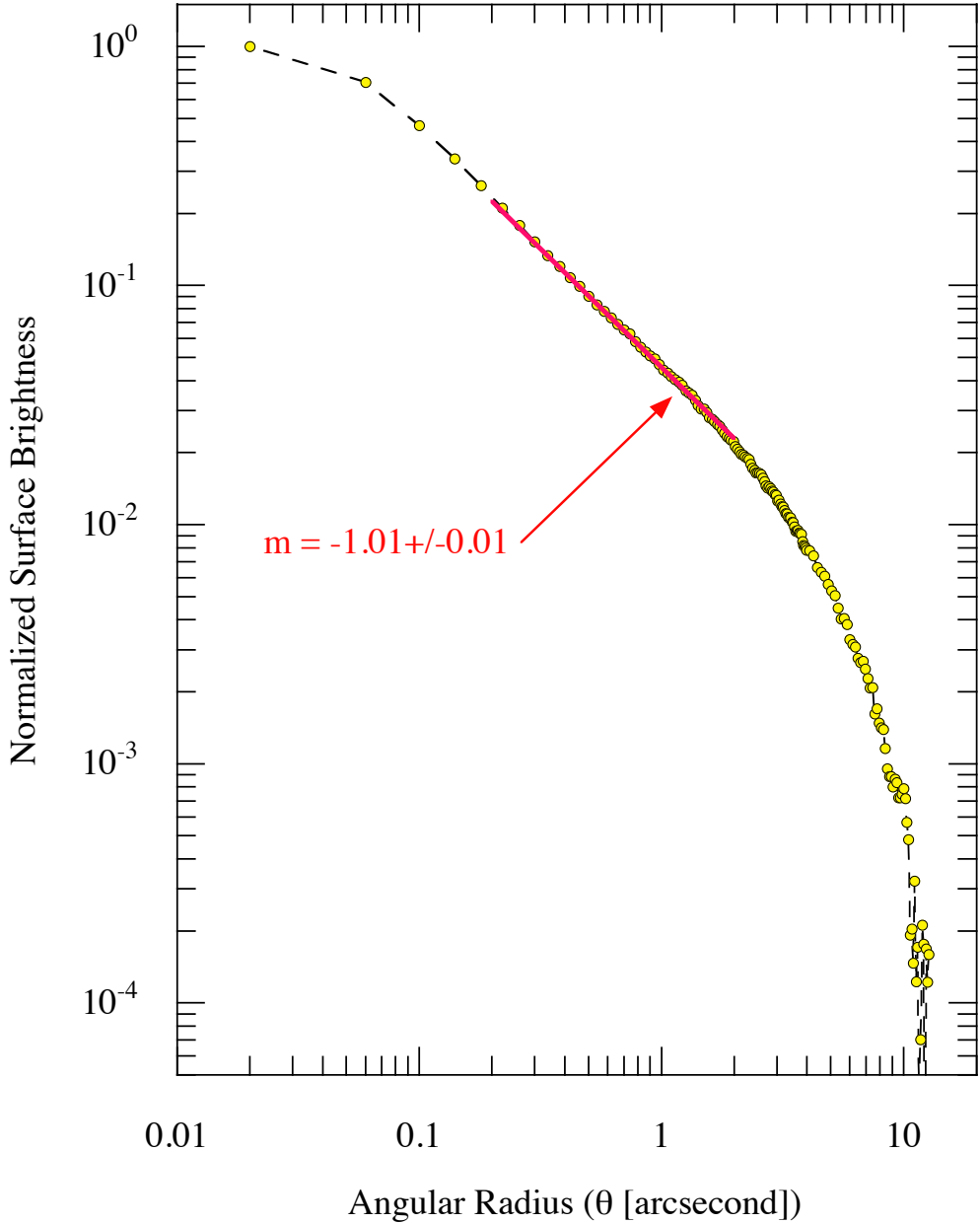}
\caption{Surface brightness profile determined in annular bins  0.04\arcsec~wide for $\theta < $ 4\arcsec~and 0.16\arcsec~wide otherwise, with sky subtraction from a surrounding annulus extending from 20\arcsec~to 22\arcsec.  The red  line segment shows a slope of m = -1 and accurately fits the profile in the range 0.2\arcsec~$\le \theta \le$ 2.0\arcsec.   \label{profile}}
\end{figure}

The  surface brightness profile, $\Sigma(\theta)$, computed using circular azimuthal averaging, consists of three parts (Figure \ref{profile}).  In the central region ($\theta <$ 0.2\arcsec)  the profile is affected by convolution with the point-spread function of HST (see below).  In the middle region, radius range 0.2\arcsec~$\le \theta \le$ 2\arcsec, a least-squares power-law fit to the profile, $\Sigma(\theta) \propto \theta^{m}$, gives the logarithmic gradient  $m$ = -1.01$\pm$0.01.  This is close to the canonical value, $m$ = -1, expected for a coma expanding in steady-state and is very different from the value, $m$ = -3/2, expected from the action of solar radiation pressure on a steady-state coma (Jewitt and Meech 1987b).   The absence of evidence for the effects of radiation pressure is strengthened by the lack of the familiar bow-wave shaped ``nose'' of the coma in the sunward direction (Figure \ref{image}).  

In the outer region ($\theta >$ 2\arcsec, corresponding to 23,000 km at the comet) the gradient progressively steepens, with $\Sigma$ reaching 1\% of the core value at $\theta$ = 3.6\arcsec~and 0.1\% at $\theta$ = 9\arcsec.  The steepening is azimuthally symmetric in the plane of the sky and suggests an edge to the coma, rather than the effects of deflection of dust particle trajectories by radiation pressure.  Both the symmetric coma and the steep-edged  profile distinguish K2 from many comets reported in the literature (Jewitt and Meech 1987b, Meech et al.~2009, S{\'a}rneczky et al.~2016) but do resemble the sharply truncated profile of long-period comet C/1980 E1 (Bowell) (Jewitt et al.~1982).  Because the sky noise grows substantially at larger radii, we take $\theta$ = 9\arcsec~as the best estimate of the radius of the coma (corresponding to linear radius $r_c$ = 1.0$\times$10$^8$ m at $\Delta$ = 15.824 AU).  

We attempted to isolate the nucleus of K2 using the surface brightness profile.  Photometry using the smallest practical aperture, 0.2\arcsec~in radius, with sky subtraction from a contiguous annulus of outer radius 0.28\arcsec, yielded apparent magnitude V = 23.27$\pm$0.01 ($H$ = 11.02$\pm$0.01, corresponding to radius $r_n$ = 22 km with albedo $p_V$ = 0.04).  However, given that simple aperture photometry blends light scattered from the nucleus and near-nucleus coma, this must represent a strong upper limit to the possible radius of the nucleus.   We sought to better  isolate the nucleus by fitting and subtracting a model of the two-dimensional surface brightness of the coma.  The model, based on Lamy et al~(2004) and references therein, fits a power-law relation along multiple azimuths within an annulus and then extrapolates the fit to zero radius and convolves with the point-spread function (PSF) to model the inner coma.  We experimented with a range of fitting radii from 0.2\arcsec~to 0.4\arcsec~at the inner edge  to outer radii from 1.0\arcsec~to 2.0\arcsec~(i.e. the power law portion of the profile in Figure \ref{profile}).  The PSF was obtained from the on-line TinyTim routine (Biretta 2014).
Our best estimate is that the nucleus has magnitude $V >$ 25.2, corresponding to a nucleus absolute magnitude $H >$ 12.9. With $p_V$ = 0.04, Equation (\ref{area}) gives cross-section $C_n <$  260 km$^2$.  The radius of an equal-area circle is $r_n <$ 9 km, which is our best estimate of the size of the nucleus.

The simplest interpretation of the data suggested by the circular isophotes (Figure \ref{image}) and by the $m$ = -1 surface brightness gradient (Figure \ref{profile}) is that the coma of K2 is in steady-state expansion (Jewitt and Meech 1987b).   If we assume that the earliest observations of activity in K2 (in 2013) correspond to the  release of the dust particles, then the timescale of the expansion is $t \sim$ 10$^8$ s, and the mean velocity of the particles over this interval is $v = r_c/t$, or $v$ = 1 m s$^{-1}$.   This  is a factor $\sim$100 times smaller than the thermal velocity in gas at the distance of K2.  In comets closer to the Sun, low dust speeds are usually associated with larger particles (which are poorly dynamically coupled to the outflowing gas; Agarwal et al.~2016).  

The absence of evidence for radiation pressure acceleration independently suggests that the coma dust particles must be large.   The radiation pressure induced acceleration is $\beta g_{\odot}(1)/r_H^2$, where $r_H$ is the heliocentric distance expressed in AU, $g_{\odot}(1)$ = 0.006 m s$^{-2}$ is the gravitational acceleration to the Sun at $r_H$ = 1 AU and $\beta$ is the dimensionless radiation pressure efficiency, approximately equal to the inverse of the grain radius expressed in microns, $\beta \sim a_{\mu m}^{-1}$  (Bohren and Huffman 1983).  Then the distance of deflection is just $\ell = \beta g_{\odot}(1)  t^2 /(2r_H^2)$, where $t$ is the time of flight.    Substituting $t \sim$ 10$^8$ s we find that micron-sized particles ($\beta$ = 1) should travel a distance $\ell \sim$ 1 AU, and should occupy a long tail, which is not observed.  The interpretation is compromised somewhat by the viewing geometry (phase angle $\alpha$ = 4\degr~for the HST observation), since we can measure the dust distribution only in the plane of the sky.  Assuming dust accelerated in the antisolar direction we substitute $\ell <  r_c /(\sin(\alpha))$, from which we obtain $\beta <$ 10$^{-2}$, corresponding to effective dust radius $a_{\mu m} \gtrsim$ 0.1 mm.  This large particle size is qualitatively consistent with a model  in which small particles are retained by cohesive forces (Gundlach et al.~2015), but is quantitatively inconsistent because their model predicts that no dust of any size can be ejected beyond $r_H \sim$ 5 AU, clearly in violation of our observations.

The mass of the dust particles, considered as spheres of density $\rho$, is $M_d = 4 \rho a C_d/3$.  Substituting $a$ = 0.1 mm, $\rho$ = 500 kg m$^{-3}$, and $C_d$ = 9.3$\times$10$^4$ km$^2$ (Table \ref{photometry}), we find $M_d \sim$ 6$\times$10$^{9}$ kg equal to roughly 10$^{-6}$ of the nucleus mass, if $r_n$ = 9 km.  If spread uniformly over the surface of a spherical nucleus of radius $r_n$ and  density $\rho$, they would form a layer of thickness $\Delta r = a C_d/(3 \pi r_n^2)$.   For example, with $r_n$ = 9 km, we find $\Delta r$ = 1 cm, comparable to the likely diurnal thermal skin depth.  If released steadily over the $t \sim$ 10$^8$ s active lifetime of K2, the average mass loss rate is $dM/dt \sim$60 kg s$^{-1}$, a remarkable value for a comet beyond Saturn.

\begin{figure}
\epsscale{.9}
\plotone{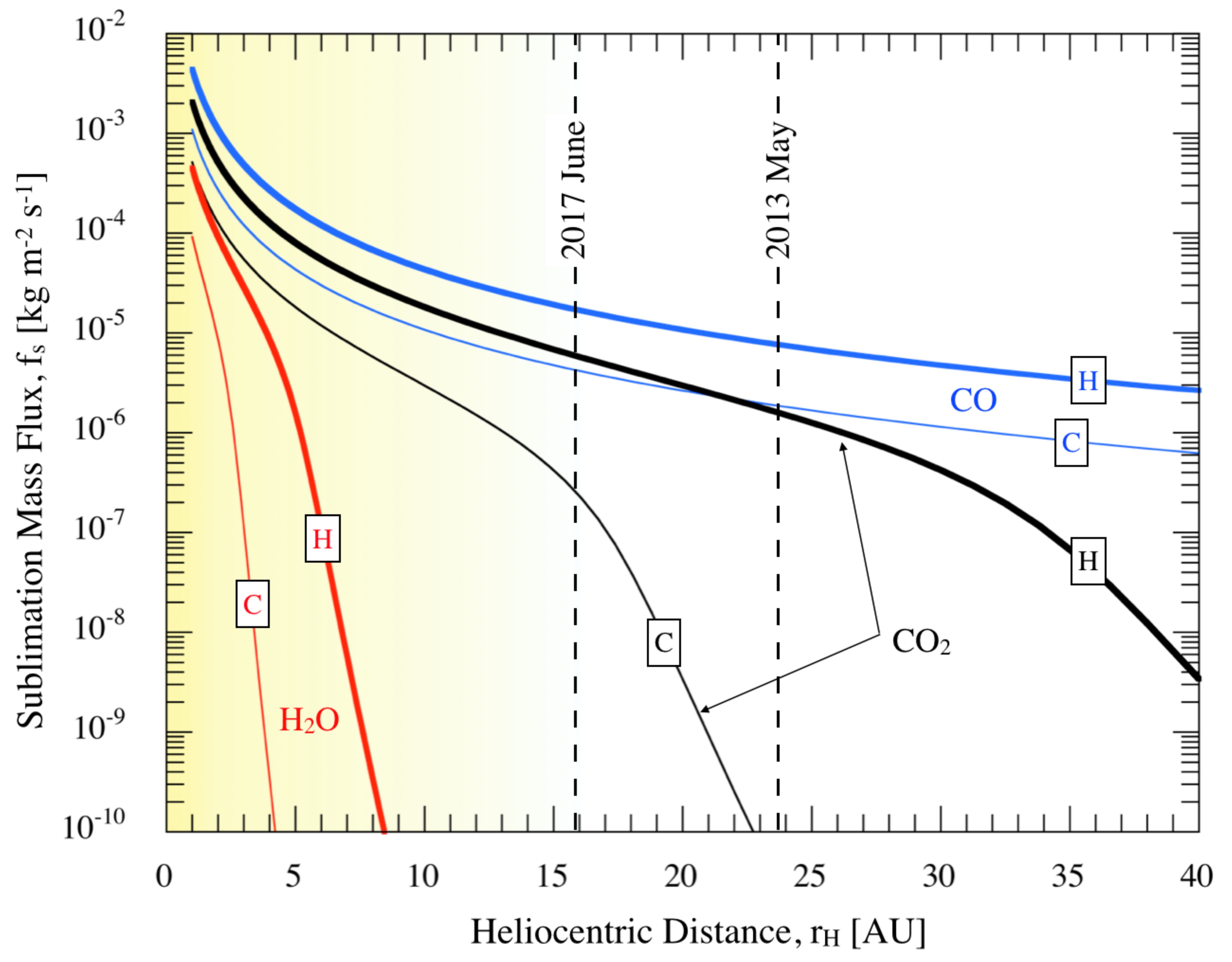}
\caption{Specific mass sublimation rates as a function of heliocentric distance for three ices of (red) H$_2$O, (black) CO$_2$ and (blue) CO, from Equation (\ref{sublimation}).  The curves for O$_2$ and N$_2$ are similar to that for CO, and are not plotted for clarity. For each ice, the sublimation rates for the minimum (labeled C for ``cold'') and maximum (H for ``hot'')  possible temperatures are indicated by thin and thick lines, respectively.  The shaded region shows the heliocentric distances where crystallization is possible.  The heliocentric distances at which the CFHT and HST observations were taken are marked by vertical dashed lines.  \label{ices}}
\end{figure}

We solved the radiative thermal equilibrium equation for H$_2$O, CO$_2$ and CO ices,

\begin{equation}
\frac{L_{\odot}}{4 \pi r_H^2}(1-A)  = \chi\left[ \epsilon \sigma T^4 + L(T) f_s(T)\right]
\label{sublimation}
\end{equation}

\noindent in which $L_{\odot}$ (W) is the luminosity of the Sun, $r_H$ (m) is the heliocentric distance, $A$  is the Bond albedo,  $\epsilon$  is the emissivity of the body, $\sigma$ (W m$^{-2}$ K$^{-4}$) is the Stefan-Boltzmann constant and $L(T)$ (J kg$^{-1}$) is the latent heat of sublimation of the relevant ice at temperature, $T$ (K).  Dimensionless parameter $\chi$ characterizes the way in which incident heat is deposited over the surface. To explore the range of possible solutions, we consider the limiting cases $\chi$ = 1, which describes the highest possible temperatures found at the subsolar point on a non-rotating nucleus and $\chi$ = 4, corresponding to the  temperature of an isothermal sphere.  The quantity $ f_s$ (kg m$^{-2}$ s$^{-1}$) is the sought-after mass flux of sublimated ice.   The two terms on the right represent power radiated from the surface into space and power used to sublimate ice, while heat conduction into the nucleus interior is neglected.  The term on the left represents power absorbed from the Sun.  To solve Equation (\ref{sublimation}) we used ice thermodynamic parameters from Brown and Ziegler (1980) and Washburn (1926) and assumed $A$ = 0.04, $\epsilon$ = 0.9.  Solutions to Equation (\ref{sublimation}) show that  sublimation of water ice cannot drive the observed activity of K2 (Figure \ref{ices}).   

The timescale for the crystallization of amorphous ice, $\tau_{CR}$ (years), at temperature, $T$, is 

\begin{equation}
\tau_{CR} = 3\times 10^{-21} \exp\left(\frac{E_A}{k T}\right)
\label{CR}
\end{equation}

\noindent where $k$ is Boltzmann's constant and  $E_A/k$ = 5370 K (Schmitt et al.~1989).  The comet is infalling on a nearly radial orbit. Accordingly, we estimate the critical distance for the crystallization of surface ice by comparing $\tau_{CR}$ with the free-fall timescale, $\tau_{ff}$, given by 

\begin{equation}
\tau_{ff} = \left(\frac{r_H^3}{2 G M_{\odot}}\right)^{1/2},
\label{ff}
\end{equation}

\noindent in which $G$ is the gravitational constant and $M_{\odot}$ is the mass of the Sun.   We reason that crystallization will occur when $\tau_{CR} \ll \tau_{ff}$.  Solving  Equations (\ref{sublimation}), (\ref{CR}) and (\ref{ff}) numerically, we find that the inequality is satisfied for the high temperature ($\chi$ = 1 in Equation \ref{sublimation}) limit at $r_H \le$ 12.5 AU and, for the low temperature limit ($\chi$ = 4) at $r_H \le $ 6.0 AU.  Both distances are small compared to the heliocentric distance of K2 in the observations discussed here, indicating that crystallization has not occurred.  A slightly higher critical distance ($r_H$ = 16 AU) was found for Centaurs  by Guilbert-Lepoutre (2012) but this reflects the much longer dynamical timescales for Centaur heating (10 Myr in her integrations compared with $\tau_{ff} \sim$ 5 yr here) and is inapplicable to the  plunge orbit of K2.
 
We thus conclude that, unlike other inbound comets that have been studied (Meech et al.~2009, 2017), K2 lies beyond \textit{both} the water ice sublimation ($r_H \lesssim$ 5 AU)  and crystallization  ($r_H \le$ 12.5 AU) zones,  ruling out these processes as sources of the observed activity.  Instead, more volatile ices might drive the activity, as shown in Figure (\ref{ices}). For example, at the sub-solar point ($\chi$ = 1) and at 23.8 AU (2013 May), carbon dioxide (latent heat of vaporization $L$ = 6$\times$10$^5$ J kg$^{-1}$) would sublimate at $f_s$ = 2$\times$10$^{-6}$ kg m$^{-2}$ s$^{-1}$, while the more volatile carbon monoxide, oxygen and nitrogen ices (all have $L \sim$ 2$\times$10$^5$ J kg$^{-1}$) would sublimate at $f_s$ = 8$\times$10$^{-6}$ kg m$^{-2}$ s$^{-1}$ (Figure \ref{ices}).  An exposed ice patch having area $f_s^{-1} (dM/dt) \sim$ 1 to 10 km$^2$ would be sufficient to supply the mass loss rate.  On a non-rotating, 9 km radius spherical nucleus with density 500 kg m$^{-3}$, these sublimation fluxes are just sufficient to expel particles (of equal density) against gravity provided their radii are $a \lesssim$ 0.2 mm and $a \lesssim$ 0.7 mm, respectively (Jewitt et al.~2014), consistent with the $a \gtrsim$ 0.1 mm particle size inferred from the absence of radiation pressure deflection of the coma.  It is thus plausible that the dust  particles in the coma of K2 were launched by gas drag from sublimating supervolatile ices,  compatible with limited laboratory evidence  for low temperature (40 K to 60 K) sublimation of supervolatile coatings on grains (Bar-Nun et al.~2007). Supervolatiles can also explain the measured brightening in the absolute magnitude from H = 8.6$\pm$0.5 at 23.8 AU to H = 7.2 at 15.9 AU.  The ratio of equilibrium sublimation rates for CO, O$_2$ and N$_2$ at these distances is 2.2:1, corresponding to a brightening by 0.9 magnitudes, while for CO$_2$ the ratio is  3.8 (1.4 magnitudes),  both comparing favorably with the observed 1.4$\pm$0.5 magnitudes within the uncertainties.  In addition, of the explanations listed in the introduction, supervolatile sublimation is the only one naturally providing sustained (as opposed to burst-like) activity over this extreme distance range. We conclude that supervolatile sublimation is the likely source of the activity in K2.

The emerging picture of K2 is of a $<$9 km radius nucleus ejecting submillimeter-sized particles at low velocities over periods of years,  driven by the sublimation of supervolatile ices. The presence of supervolatile ices in the near-surface regions of the comet is consistent with its long period orbit and probably dynamically new nature.  In these regards K2 resembles the inbound long-period comet C/1980 E1 (Bowell) which,  at $r_H <$ 7 AU, displayed a spherical coma  with a nearly parallel-sided tail  consisting of 0.3 to 1 millimeter-sized particles of considerable age (Sekanina 1982).  The coma expanded linearly at speed 0.9$\pm$0.2 m s$^{-1}$ in observations over the heliocentric distance range $\sim$5.0 AU to $\sim$3.5 AU (Jewitt 1984), similar to the expansion rate of K2.    Outgassing of OH from C/1980 E1 only became strong near $r_H$ = 4.6 AU, apparently caused by sublimation of icy grains in the coma, while the nucleus itself activated only near perihelion  (at 3.36 AU, A'Hearn et al.~1984).  
The early detection of K2 (at $\sim$24 AU vs.~$\sim$7 AU for C/1980 E1) will allow a much richer investigation of the behavior of a long period comet entering the planetary region, leading to an improved understanding of the processes occurring when warming up from Oort cloud temperatures.

\acknowledgments
We thank Pedro Lacerda  and the anonymous referee for comments.  Based on observations made under GO 14939 with the NASA/ESA Hubble Space Telescope, obtained at the Space Telescope Science Institute,  operated by the Association of Universities for Research in Astronomy, Inc., under NASA contract NAS 5-26555.   This research used the facilities of the Canadian Astronomy Data Centre, National Research Council of Canada with the support of the Canadian Space Agency.  DJ appreciates support from NASA's Solar System Observations program.



{\it Facilities:}  \facility{Hubble Space Telescope}.




\clearpage


\begin{deluxetable}{llcccccr}
\tablecaption{Observing Geometry 
\label{geometry}}
\tablewidth{0pt}
\tablehead{ \colhead{Object} & \colhead{UT Date and Time}   & $r_H$\tablenotemark{a} & $\Delta$\tablenotemark{b} & \colhead{$\alpha$\tablenotemark{c}} & $\theta_{\odot}$\tablenotemark{d}  & $\theta_{-V}$\tablenotemark{e}  & $\delta_{\oplus}$\tablenotemark{f} }
\startdata

CFHT& 2013 May 12 14:06 - 14:28 &  23.742 & 23.765  & 2.4 & 215.9 & 354.2   & -1.44  \\
HST & 2017 Jun 27  ~20:08 - 20:48 & 15.874 & 15.816 & 3.7 & 167.4 & 357.1 & 0.52  \\

\enddata


\tablenotetext{a}{Heliocentric distance, in AU}
\tablenotetext{b}{Geocentric distance, in AU}
\tablenotetext{c}{Phase angle, in degrees}
\tablenotetext{d}{Position angle of anti-solar direction, in degrees}
\tablenotetext{e}{Position angle of negative projected orbit vector, in degrees}
\tablenotetext{f}{Angle from orbital plane, in degrees}

\end{deluxetable}

\clearpage

%
%
%
%
%

\clearpage

\begin{deluxetable}{lcccccc}
\tablecaption{HST Fixed-Aperture Photometry\tablenotemark{a} 
\label{photometry}}
\tablewidth{0pt}

\tablehead{ \colhead{Quantity} & $ $ 5 & 10  & 20 & 40 & 80 & 160}
\startdata
V [mag]\tablenotemark{b} &  	21.56 & 20.79 & 20.02	& 19.33 & 18.83 & 18.66  \\
H [mag]\tablenotemark{c} &           9.41 & 8.64 & 7.87 & 7.18 & 6.68 & 6.51  \\
$C_d/1000$ [km$^2$]\tablenotemark{d} &           6.5 & 13.1 & 26.7 & 50.4 & 79.8 &  93.3  \\
\hline

\enddata

\tablenotetext{a}{Aperture radii  in units of 10$^3$ km at the comet }
\tablenotetext{b}{Mean apparent V magnitude }
\tablenotetext{c}{Absolute V magnitude calculated from Equation (\ref{abs})}
\tablenotetext{d}{Scattering cross-section $\times$ 10$^{-3}$ km$^2$, from Equation (\ref{area})}

\end{deluxetable}

\clearpage

\clearpage

\clearpage

\clearpage


\begin{thebibliography}{}

\bibitem[Agarwal et al.(2016)]{2016MNRAS.462S..78A} Agarwal, J., A'Hearn, M.~F., Vincent, J.-B., et al.\ 2016, \mnras, 462, S78 

\bibitem[A'Hearn et al.(1984)]{1984AJ.....89..579A} A'Hearn, M.~F., Schleicher, D.~G., Millis, R.~L., Feldman, P.~D., \& Thompson, D.~T.\ 1984, \aj, 89, 579 

\bibitem[Bar-Nun et al.(2007)]{2007Icar..190..655B} Bar-Nun, A., Notesco, G., \& Owen, T.\ 2007, \icarus, 190, 655 

\bibitem[Biretta(2014)]{2014wfc..rept...10B} Biretta, J.\ 2014, Space Telescope WFC Instrument Science Report,  Space Telescope Science Institute, Baltimore, Maryland.

\bibitem[Bohren \& Huffman(1983)]{1983asls.book.....B} Bohren, C.~F., \& Huffman, D.~R.\ 1983, Absorption and scattering of light by small particles, New York: Wiley, 1983,  

\bibitem[Brown\& Ziegler(1980)]{1980BrownAdvCryoEng}  Brown, G. and Ziegler W. (1980).  Adv.~Cryog.~Eng.~25, 662-670.

\bibitem[Dones et al.(2015)]{2015SSRv..197..191D} Dones, L., Brasser, R., Kaib, N., \& Rickman, H.\ 2015, \ssr, 197, 191 

\bibitem[Donn \& Urey(1956)]{1956ApJ...123..339D} Donn, B., \& Urey, H.~C.\ 1956, \apj, 123, 339 



\bibitem[Guilbert-Lepoutre(2012)]{2012AJ....144...97G} Guilbert-Lepoutre, A.\ 2012, \aj, 144, 97 

\bibitem[Gundlach et al.(2015)]{2015A&A...583A..12G} Gundlach, B., Blum, J., Keller, H.~U., \& Skorov, Y.~V.\ 2015, \aap, 583, A12 





\bibitem[Jewitt(1984)]{1984Icar...60..373J} Jewitt, D.\ 1984, \icarus, 60, 373 

\bibitem[Jewitt(2015)]{2015AJ....150..201J} Jewitt, D.\ 2015, \aj, 150, 201 

\bibitem[Jewitt et al.(1982)]{1982AJ.....87.1854J} Jewitt, D.~C., Soifer, B.~T., Neugebauer, G., Matthews, K., \& Danielson, G.~E.\ 1982, \aj, 87, 1854 


\bibitem[Jewitt \& Meech(1987)]{1987AJ.....93.1542J} Jewitt, D., \& Meech, K.\ 1987a, \aj, 93, 1542 

\bibitem[Jewitt \& Meech(1987)]{1987ApJ...317..992J} Jewitt, D.~C., \& Meech, K.~J.\ 1987b, \apj, 317, 992 

\bibitem[Jewitt et al.(2014)]{2014AJ....147..117J} Jewitt, D., Ishiguro, M., Weaver, H., et al.\ 2014, \aj, 147, 117 
%
Title:	


\bibitem[Kulyk et al.(2016)]{2016Icar..271..314K} Kulyk, I., Korsun, P., Rousselot, P., Afanasiev, V., \& Ivanova, O.\ 2016, \icarus, 271, 314 

\bibitem[Lamy et al.(2004)]{2004come.book..223L} Lamy, P.~L., Toth, I., Fernandez, Y.~R., \& Weaver, H.~A.\ 2004, Comets II, 223 

\bibitem[Levison(1996)]{1996ASPC..107..173L} Levison, H.~F.\ 1996, Completing the Inventory of the Solar System, 107, 173 

\bibitem[Meech \& Jewitt(1987)]{1987A&A...187..585M} Meech, K.~J., \& Jewitt, D.~C.\ 1987, \aap, 187,  

\bibitem[Meech et al.(2009)]{2009Icar..201..719M} Meech, K.~J., Pittichov{\'a}, J., Bar-Nun, A., et al.\ 2009, \icarus, 201, 719 

\bibitem[Meech et al.(2017)]{2017AJ....153..206M} Meech, K.~J., Schambeau, C.~A., Sorli, K., et al.\ 2017, \aj, 153, 206 


\bibitem[Miles(2016)]{2016Icar..272..356M} Miles, R.\ 2016, \icarus, 272, 356 

\bibitem[Prialnik \& Bar-Nun(1990)]{1990ApJ...363..274P} Prialnik, D., \& Bar-Nun, A.\ 1990, \apj, 363, 274 


\bibitem[Rettig et al.(1992)]{1992ApJ...398..293R} Rettig, T.~W., Tegler, S.~C., Pasto, D.~J., \& Mumma, M.~J.\ 1992, \apj, 398, 293 


\bibitem[Rickman(2014)]{2014M&PS...49....8R} Rickman, H.\ 2014, Meteoritics and Planetary Science, 49, 8 

\bibitem[S{\'a}rneczky et al.(2016)]{2016AJ....152..220S} S{\'a}rneczky, K., Szab{\'o}, G.~M., Cs{\'a}k, B., et al.\ 2016, \aj, 152, 220 

\bibitem[Schmitt et al.(1989)]{1989ESASP.302...65S} Schmitt, B., Espinasse, S., Grim, R.~J.~A., Greenberg, J.~M., \& Klinger, J.\ 1989, in ESA Physics and Mechanics of Cometary Materials, 302, 65

\bibitem[Sekanina(1973)]{1973NASSP.319..199S} Sekanina, Z.\ 1973, NASA Special Publication, 319, 199 

\bibitem[Sekanina(1982)]{1982AJ.....87..161S} Sekanina, Z.\ 1982, \aj, 87, 161 


\bibitem[Solontoi et al.(2012)]{2012Icar..218..571S} Solontoi, M., Ivezi{\'c}, {\v Z}., Juri{\'c}, M., et al.\ 2012, \icarus, 218, 571 


\bibitem[Wainscoat et al.(2017)]{2017CBET.4393....1W} Wainscoat, R.~J., Wells, L., Micheli, M., \& Sato, H.\ 2017, Central Bureau Electronic Telegrams, 4393,  

\bibitem[Washburn(1926)]{1926washburn} Washburn, E.\ 1926, International Critical Tables of Numerical data, Physics, Chemistry and Technology, Vol. 3 (New York: McGraw-Hill).

\bibitem[Whipple(1950)]{1950ApJ...111..375W} Whipple, F.~L.\ 1950, \apj, 111, 375 


\bibitem[Womack et al.(2017)]{2017PASP..129c1001W} Womack, M., Sarid, G., \& Wierzchos, K.\ 2017, \pasp, 129, 031001 

\end{thebibliography}
\end{document}